\documentclass[%
 reprint,
 amsmath,amssymb,
 aps,
]{revtex4-2}

\usepackage{xcolor}
\usepackage{graphicx}
\usepackage{dcolumn}
\usepackage{bm}
\usepackage{hyperref}

\usepackage{amsmath}
\usepackage{float}
\usepackage{amsfonts}
\usepackage{soul}
\usepackage{afterpage}
\usepackage{placeins}

\hypersetup{
    colorlinks=true,
    linkcolor=blue,
    filecolor=black,
    anchorcolor=black,
    citecolor=blue,
    filecolor=black,
    urlcolor=blue
    }

\newcommand{\im}{{\rm i}}

\begin{document}

\preprint{APS/123-QED}

\title{Propagation-invariant optical meron lattices} \author{David Marco\textsuperscript{1, 2}}
\author{Isael Herrera\textsuperscript{1}}
\author{Sophie Brasselet\textsuperscript{1}}
\author{Miguel A. Alonso\textsuperscript{1, 3, 4, 5,}}
 \email{miguel.alonso@fresnel.fr}
\affiliation{%
  \textsuperscript{1}Aix Marseille Univ., CNRS, Centrale Med, Institut Fresnel, UMR 7249, 13397 Marseille Cedex 20, France\\
 \textsuperscript{2}Instituto de Bioingeniería, Universidad Miguel Hernández de Elche, 03202 Elche, Spain\\
 \textsuperscript{3}The Institute of Optics, University of Rochester, Rochester, NY14627, USA\\
 \textsuperscript{4}Center for Coherence and Quantum Optics, University of Rochester, Rochester, NY14627, USA\\
 \textsuperscript{5}Laboratory for Laser Energetics, University of Rochester, Rochester, NY14627, USA
}%

\begin{abstract}
We introduce and produce experimentally optical beams exhibiting periodic skyrmionic polarization lattices at each transverse plane of propagation. These textures are meron lattices formed by tiles mapping hemispheres of the Poincaré sphere. All presented fields are combinations of a small number of plane waves. Firstly, we propose square lattices with a Skyrme density (the Jacobian of the mapping between the Poincaré sphere and physical space) that oscillates in sign but whose intensity distribution is constant. Secondly, we present triangular lattices preserving the Skyrme density's sign. Both lattices are invariant under propagation. Finally, we introduce a family of lattices with uniform Skyrme density sign, composed of square tiles that map to the same hemisphere of the Poincaré sphere. In these lattices, the polarization state undergoes a uniform local periodic rotation during propagation, thus preserving the texture's Skyrme density distribution.
\end{abstract}

\maketitle

\section{Introduction}
   
    Skyrmionic textures emerge in the spatial distribution of a vector field within a region covering all possible local field states, where each state is represented by a point on a spherical parameter space \cite{Skyrme, magnetic_skyrmions_review}. Two-dimensional skyrmions arise in regions where a vector field spans a 2-sphere without reversing the sign of the Jacobian of the transformation, known as the Skyrme density, $\rho_\mathrm{S}$. The Skyrme number, $N_\mathrm{S}$, is the integral of $\rho_\mathrm{S}$ within the Skyrmiom's region divided by the sphere's area. This is an integer that represents the number of times the sphere has been covered, with its sign denoting the sense. Another 2D topological structure is the meron, representing a region in space where only one hemisphere of the spherical parameter space is mapped \cite{beyond_skyrmions}. Beyond 2D, hopfions emerge in 3D regions that map a full 3D spherical parameter space, realizing the Hopf fibration \cite{magnetic_hopfions,optical_hopfion}.

    Multiple physical systems have been found to present skyrmionic textures, including magnetic materials \cite{magnetic_skyrmions_review, beyond_skyrmions, meron_lattices_magnetic} giving rise to skyrmion lattices \cite{magnetic_skyrmions_review}, meron lattices \cite{meron_lattices_magnetic}, and hopfions \cite{magnetic_hopfions}. Additionally, skyrmionic distributions have been investigated in superfluids \cite{meron_lattice_superfluid, meron_lattice_superfluid_book}, sound waves \cite{acoustic_skyrmions, Muelas_soundwaves}, liquid crystals \cite{LC_skyrmions}, and water waves \cite{skyrmions_water}. 
    
    In recent years, skyrmionic textures have been discovered in optical fields \cite{review_optical_skyrmions_Shen,first_optical_skyrmions,skyrmion_spin_evanescent,skyrmions_Rodrigo_Pisanty,optical__merons_PRL,optical__merons_Zhang,optical_plasmonic_merons,spin_merons_polygons, Airy_beams_merons,Poincare_skyrmions,paraxial_skyrmionic_beams,Shen_bimerons, skyrmions_lens,lemon_fields,cartography_skyrmions}. In nonparaxial optical fields, hexagonal lattices of skyrmions emerge from the instantaneous local orientation of the electric field in evanescent waves \cite{first_optical_skyrmions}. Optical skyrmions are also found in the distribution of the local normalized spin density vector \cite{skyrmion_spin_evanescent,skyrmions_Rodrigo_Pisanty}. Additionally, spin optical meron lattices have been implemented \cite{optical__merons_PRL,optical__merons_Zhang,optical_plasmonic_merons, spin_merons_polygons}. Beyond 2D textures, periodic skyrmionic distributions spanning a 4-sphere embedded in a 5D ambient space were found within a spatio-temporal region of a monochromatic field where all possible nonparaxial polarization ellipses are achieved \cite{F3DP_fields}.

    Two-dimensional skyrmionic distributions spanning the Poincaré sphere, known as Stokes textures \cite{review_optical_skyrmions_Shen}, were identified in the polarization state distribution of paraxial optical fields containing all possible values of the normalized Stokes vector within a region of a beam's transverse plane. Full Poincaré beams \cite{full_Poincare_beams} are paraxial optical skyrmions \cite{Poincare_skyrmions, paraxial_skyrmionic_beams} performing a stereographic projection of the Poincaré sphere at every transverse plane. The implementation of Stokes textures using spatial light modulators \cite{cartography_skyrmions} or other diffractive optical elements \cite{skyrmions_lens} is relatively straightforward, making them ideal for exploring distributions beyond skyrmions, such as bimerons, skyrmioniums, multiskyrmions, or multimerons \cite{Shen_bimerons, skyrmions_lens}. Moreover, recent discoveries include optical hopfions within the volume of propagation of a paraxial field \cite{optical_hopfion}.

    Stokes meron lattices emerge in polarization lattices comprising tiles spanning hemispheres of the Poincaré sphere \cite{lemon_fields,stars_fields,cartography_skyrmions}. Stokes lattices often exhibit oscillations in the Skyrme density sign \cite{lemon_fields,stars_fields}, $\mathrm{sgn}(\rho_\mathrm{S})$, as observed in optical spin meron lattices \cite{optical__merons_PRL,optical__merons_Zhang,optical_plasmonic_merons, spin_merons_polygons}, similar to the behavior observed in magnetic meron lattices \cite{meron_lattices_magnetic}. Such oscillations may lead in practice to $N_\mathrm{S} \approx 0$ within an area containing several merons, similar to certain nonskyrmionic regions. On the other hand, textures preserving $\mathrm{sgn}(\rho_\mathrm{S})$ have been observed in \textsuperscript{3}He-A superfluid meron lattices \cite{meron_lattice_superfluid, meron_lattice_superfluid_book} and sound waves \cite{Muelas_soundwaves}. A periodic texture preserving $\mathrm{sgn}(\rho_\mathrm{S})$ wraps over a spherical parameter space in the same direction periodically along any spatial direction. Recently, textures based on conformal cartographic maps with uniform $\mathrm{sgn}(\rho_\mathrm{S})$ were introduced \cite{cartography_skyrmions}, mapping either the sphere or a hemisphere onto every regular polygon that provides regular tessellations. When implemented as Stokes textures \cite{cartography_skyrmions}, they render fields where the texture is strongly perturbed under propagation. In contrast, propagation-invariant Stokes meron lattices in the literature \cite{lemon_fields, stars_fields} do not preserve $\mathrm{sgn}(\rho_\mathrm{S})$.
 
    Here, we introduce optical beams exhibiting periodic skyrmionic textures in their polarization state distribution at each plane perpendicular to the direction of propagation, some of them preserving the sign of the Skyrme density. Experimental approximations to these beams showing the desired features are also presented. The first fields we discuss are propagation-invariant lattices of square merons, created by superposing four plane waves. In this case, $\mathrm{sgn}(\rho_\mathrm{S})$ varies spatially while the intensity remains constant across the entire space. Additionally, we introduce a more unusual family of fields: triangular meron lattices that not only exhibit invariance under propagation but also maintain a uniform $\mathrm{sgn}(\rho_\mathrm{S})$. These fields are composed of six plane waves. 
   Finally, we propose an even more exotic family of fields exhibiting lattices of square merons where, remarkably, each meron maps exclusively the same Poincaré hemisphere (either northern or southern), while at the same time preserving $\mathrm{sgn}(\rho_\mathrm{S})$. That is, the entire lattice covers just a single hemisphere, wrapping it repeatedly in the same sense over the transverse plane. 
    These fields are composed of a mixture of eight plane waves. The only transformation that this polarization state pattern undergoes as it propagates is a uniform local rotation of the polarization ellipses; 
    both the intensity and $\rho_\mathrm{S}$ distributions are preserved under propagation.

\section{Paraxial optical meron lattices}

    In this section, we describe the fundamental characteristics of meron lattices that emerge in the polarization distribution of paraxial monochromatic optical fields. The electric field vector of monochromatic light oscillates describing an elliptical path over time. In the paraxial regime, this path is predominantly confined to the plane perpendicular to the main propagation direction. The shape, sense of rotation and orientation of the ellipse define the polarization state. Mathematically, the elliptical trajectory is described by $\mathrm{Re}\left[\mathbf{E} \exp{(-\im \omega t)}\right]$, where $\omega$ is the temporal angular frequency of the field, and $\mathbf{E}$ denotes the complex transverse electric field, which can depend on the spatial coordinates $(x,y)$. This field can be expressed in terms of circular left- and right-handed polarization states ($\mathbf{l},\mathbf{r}=(\mathbf{x}\pm \im \mathbf{y})/\sqrt{2}$, where $\mathbf{x}, \mathbf{y}$ are the horizontal and vertical linearly polarized states) as:
	\begin{equation}
		\mathbf{E}  = 
		E_0 \left[ \cos(\theta/2)
		\;\mathbf{l}
		+
		\sin(\theta/2) \; e^{\im \phi}
		\;\mathbf{r} \right]=E_\mathbf{l} \mathbf{l} + E_\mathbf{r} \mathbf{r},
		\label{eq:general_field}
	\end{equation}
	with $E_0$, $\phi$, and $\theta$ being functions of $(x,y)$, and $E_{\mathbf{l},\mathbf{r}}$ being the left- and right-handed circularly polarized field components. The polarization state is defined entirely by $\theta$ and $\phi$, whereas $E_0$ is a complex factor that imparts a global amplitude and phase. The angle $\theta$ sets the relative amplitude between the circular polarization components, thereby determining the eccentricity and sense of rotation (right- or left-handedness) of the polarization ellipse. On the other hand, $\phi$ represents the relative phase between $E_{\mathbf{l}}$ and $E_{\mathbf{r}}$ ($\phi=\phi_{\mathbf{r}}-\phi_{\mathbf{l}}$, where $\phi_{\mathbf{l},\mathbf{r}}$ denote the phases of $E_{\mathbf{l},\mathbf{r}}$), and it is twice the angle between the ellipse's major axis and the $x$ axis.
	
\begin{figure*}
    \centering
    \def\svgwidth{1\textwidth}
    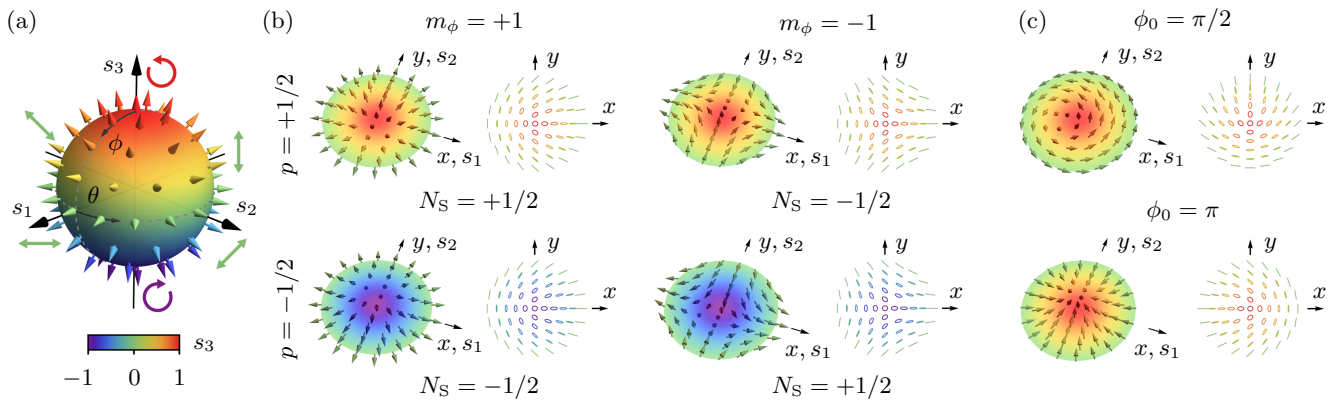\caption{\label{fig:merons}(a) The Poincaré sphere. (b) Merons classified according to their numbers $p$ and $m_\phi$. For each meron, the figure shows the normalized Stokes vector distribution and the corresponding polarization ellipse map. The orientation of the normalized Stokes vectors is depicted by choosing $s_1$, $s_2$, and $s_3$ to coincide with the Cartesian coordinates in real space $x$, $y$, and $z$, respectively. (c) Merons with  $p=+1/2$ and $m_\phi=+1$ for different initial values of $\phi$ when the polar coordinate in the plane is zero: $\phi_0=\pi/2$ (Bloch-type meron) and $\phi_0=\pi$. Note that $\phi_0=0$ for the meron with $p=+1/2$ and $m_\phi=+1$ in Fig.~\ref{fig:merons}(b) (Neel-type meron).}
    \end{figure*}

	Every polarization ellipse confined to a plane is represented as a point with spherical coordinates $\phi$ and $\theta$ on the Poincaré sphere (Fig.~\ref{fig:merons}(a)). The Cartesian coordinates of a point on the sphere are the components of the normalized Stokes vector $\mathbf{s}=(s_1,s_2,s_3)=(\sin\theta \cos\phi,\sin\theta \sin\phi,\cos\theta)$. Left-handed polarization ellipses lie on the northern hemisphere, whereas right-handed ellipses are situated on the southern hemisphere. Linear polarization states are located along the equator. The Skyrme number quantifies the number of times that $\mathbf{s}$ wraps around the Poincaré sphere within a region $\sigma$, as well as the sense of this wrapping. Mathematically, it is defined as:
	\begin{equation}
		N_\mathrm{S} =\frac{1}{4 \pi} \iint \limits_\sigma \rho_\mathrm{S} (x,y) \; \mathrm{d}x \, \mathrm{d}y,
		\label{eq:Skyrme_number1}
	\end{equation}
	where \(\rho_\mathrm{S}(x,y)\) represents the Skyrme density:
	\begin{equation}
		\rho_\mathrm{S} (x,y) = \mathbf{s}(x,y) \cdot \left[\partial_x \mathbf{s}(x,y) \times \partial_y \mathbf{s}(x,y) \right].
		\label{eq:Skyrme_density1}
	\end{equation}
	The sign of $\rho_\mathrm{S} $ depends on the sense in which the sphere is wrapped. The normalization by the area of the unit sphere in Eq.~\eqref{eq:Skyrme_number1} ensures that $N_\mathrm{S}$ within a region where the Stokes vector wraps around the entire Poincaré sphere once in the same sense is $\pm 1$. Within a meron covering the sphere once, $N_\mathrm{S}=\pm 1/2$.

    Merons can be classified according to two numbers: the polarity, $p$, and the vorticity, $m_{\phi}$. The polarity is determined by the hemisphere that is mapped to the meron: $p=\mathrm{sgn}(s_3)/2$, with $p=\pm1/2$ for the northern  and southern hemispheres, respectively. Moreover, merons contain a singularity in $\phi$ with a topological charge $m_\phi$ at the point that maps to a pole on the sphere. The number $m_{\phi}$ quantifies the number of $2\pi$-radian variations that $\phi$ experiences along a path enclosing the singular point in its proximity. For an anticlockwise (clockwise) net variation in $\phi$, $m_{\phi}>(<)0$. The Skyrme number within a meron is given by $N_\mathrm{S}= p \, m_{\phi}$ \cite{magnetic_skyrmions_review}.

    Figure \ref{fig:merons}(b) illustrates four distinct families of disk-shaped merons resulting from all combinations of $p$ and $m_{\phi}$, with $|m_{\phi}|=1$. The parameter $\phi_0$ determines the value of $\phi$ when the planar polar coordinate is zero. This value influences the form of the distribution of $\mathbf{s}$ while maintaining the numbers $m_\phi$ and $p$. For example, in Fig.~\ref{fig:merons}(b), we observe Neel-type merons for $p=+1/2$ and $m_\phi=+1$, characterized by a hedgehog-like distribution, due to $\phi_0=0$. However, by setting $\phi_0=\pi/2$ while preserving $p$ and $m_\phi$, we achieve Bloch-type merons, exhibiting a vortex-like distribution \cite{review_optical_skyrmions_Shen} (Fig.~\ref{fig:merons}(c)).
    (Note that these types are also a result of our arbitrary choice of associating the $x$ and $y$ axes in physical space with the $s_1$ and $s_2$ axes in the Stokes space, respectively.) 
	For each family of Stokes merons shown in Fig.~\ref{fig:merons}, their equivalent polarization ellipse distribution is also illustrated. When a meron maps the northern (southern) Poincaré hemisphere, it contains left (right)-handed polarization ellipses. The boundary of all the merons we study here is a curve of linear polarization, known as an L-line, which is defined by the contour where the amplitude of the $E_{\mathbf{l},\mathbf{r}}$ components is equal. A left (right)-handed polarized Stokes meron ($p=\pm1/2$) is obtained by combining a vortex in $\phi_\mathrm{r}$ ($\phi_\mathrm{l}$) having a topological charge of $m_\mathbf{r}$ ($m_\mathbf{l}$) with a region lacking a vortex in the orthogonal circular polarization component. This combination results in the aforementioned singularity in the relative phase $\phi$, with $m_\phi=\mp m_{\mathbf{l},\mathbf{r}}$. The singularity in $\phi$ gives rise to a singularity in the orientation of the polarization ellipse, commonly referred to as a C-point \cite{Freund_lattices, Dennis_singularities}, where the ellipse becomes a circle. The polarization ellipse experiences a rotation of $ \pi m_\phi/2$ rad along a path enclosing the singularity in its vicinity. This results in the well-known lemon and star patterns observed in the lines that align with the orientation of the ellipses \cite{Dennis_singularities} for $m_\phi=\pm1$, respectively. 
	
	Another type of polarization singularity arising from singularities in $\phi$ is that of V-points. Similar to C-points, the orientation of the polarization ellipse changes along a curve surrounding the singularity in close proximity. Nevertheless, for V-points, the handedness and ellipticity of the polarization ellipse surrounding the vicinity of the singularity remain constant \cite{Freund_polarization_singularities}. In a polarization lattice, either V-points or scalar vortices occur at points where vortices with the same absolute value of $m_{\mathbf{l},\mathbf{r}}$ in $E_{\mathbf{l},\mathbf{r}}$ coincide. When  $m_\mathbf{r}=-m_\mathbf{l}$, a vortex in $\phi$ with $m_\phi=2 m_\mathbf{r}$ appears (since $m_\phi=m_\mathbf{r}-m_\mathbf{l}$) leading to a V-point. Conversely, if $m_\mathbf{r}=m_\mathbf{l}$, no vortex appears in $\phi$, but a scalar vortex with a topological charge of $m_\mathbf{r}$ arises. Both types of singularities manifest as zeros in the field. These zeros are unstable, since they originate from the exact alignment of vortices in every pair of orthogonal polarization components. Any slight misalignment between these components leads to the disappearance of these zeros in practice. In contrast, a slight misalignment between $E_{\mathbf{l},\mathbf{r}}$ in a C-point will still yield a C-point, as they arise from a vortex in a polarization component with a zone lacking a vortex in the other. It is worth noting that it is possible to create unstable C-points in which the field is zero at the C-point by making coincide vortices with different absolute values of the topological charge, but such C-points do not appear in the lattices we present.

    Some of the lattices presented here have constant $\mathrm{sgn}(\rho_\mathrm{S})$ throughout all space. It can be demonstrated that all left (right)-handed C-points belonging to a Stokes skyrmionic structure with $\rho_\mathrm{S}>0$ originate from a vortex in $\phi$ with $m_\phi>(<)0$ (while in textures with $\rho_\mathrm{S}<0$, the situation is reversed) \cite{cartography_skyrmions}. As a consequence, all C-points emanating from a singularity in $\phi$ with $|m_\phi|=1$ give rise to left-handed lemon and right-handed star patterns. This statement can be easily proven for skyrmionic lattices that preserve $\mathrm{sgn}(\rho_\mathrm{S})$ composed of tiles that are polygonal versions of the merons in Fig.~\ref{fig:merons}(b). Such tiles should exclusively consist of merons with the same $N_\mathrm{S}$, i.e., polygonal versions of the merons either in the diagonal (with $N_\mathrm{S}=+1/2$) or anti-diagonal (with $N_\mathrm{S}=+1/2$) terms in Fig.~\ref{fig:merons}(b). It was shown that this result implies that periodic Stokes textures with uniform $\mathrm{sgn}(\rho_\mathrm{S})$ inevitably exhibit zeros in the electric field \cite{cartography_skyrmions}. Consequently, the zeros appearing in the fields with uniform $\mathrm{sgn}(\rho_\mathrm{S})$ that we present here cannot be eliminated.

    The main approach we follow for generating Stokes meron lattices involves seeking spatial configurations of scalar vortices obtained through the superposition of a small set of plane waves. We construct these patterns independently for each circular polarization component. By strategically setting the position and topological charges of the vortices in each circular polarization component, specific vortices in one component align with regions lacking a vortex in the other, resulting in the formation of the desired merons.

\section{Propagation-invariant textures}

    In this section, we explore combinations of plane waves with the transverse component of their wavevectors lying on a ring, thus rendering skyrmionic polarization distributions that are invariant under propagation.
    
\subsection{Square paraxial optical meron lattices} \label{sec:square_field}

   The first texture we present is a square meron lattice with uniform intensity. It is derived by first expressing $\mathbf{E}$ in Eq.~\eqref{eq:general_field} in the $\mathbf{x}$, $\mathbf{y}$ basis as $\mathbf{E} = E'_0 \left[\cos(\alpha/2) e^{-\im \delta/2} \mathbf{x} + \sin(\alpha/2) e^{\im \delta/2} \mathbf{y}\right]$, where $E'_0$, $\alpha$, $\delta$ may vary as functions of $(x,y)$. The polarization state is now determined by the parameters $\alpha$ and $\delta$, whereas $E'_0$ serves as a global amplitude and phase function. Note that the parameters $\alpha$ and $\delta$ represent spherical angles with respect to the $s_1$ axis, in contrast to $\theta$ and $\phi$, which denote spherical angles with respect to the $s_3$ axis. As a result, we have $\mathbf{s}=(\cos\alpha, \sin\alpha \cos\delta,\sin\alpha \sin\delta)$.

\begin{figure}
    \centering
    \def\svgwidth{0.48\textwidth}
    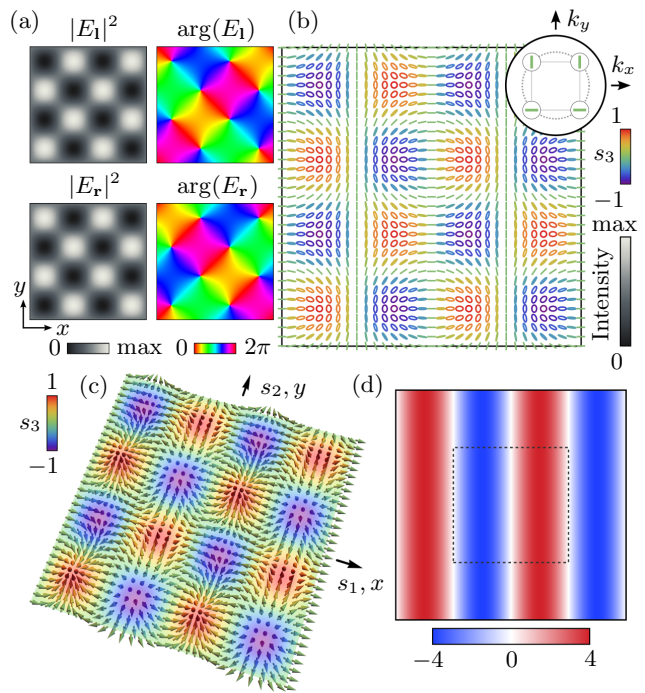\caption{\label{fig:square} Propagation-invariant square optical meron lattice with a uniform intensity distribution. (a) Intensity and phase of the circular polarization components. (b) Polarization ellipse distribution, and angular spectrum of the field (inset) where each dot represents a diffraction order associated with a plane wave whose polarization state is depicted. (c) Normalized Stokes vector distribution and (d) Skyrme density distribution divided by $k^2_\perp$. The unit cell of the texture is enclosed within dashed lines.}
    \end{figure}

    A field with a uniform intensity distribution can be achieved by making $E'_0$ constant, chosen here for simplicity as $E'_0=1$. Then, to obtain regions spanning the Poincaré sphere, we assign the angles $\alpha$ and $\delta$ in $\mathbf{E}$ to be functions that vary proportionally with the $(x,y)$ coordinates. Specifically, we assign $\alpha\rightarrow 2 \tilde{x}$ and $\delta\rightarrow 2 \tilde{y}$, where $\tilde{x} = k_\perp x$ and $\tilde{y} = k_\perp y$, with $k_\perp$ being a positive constant. This results in the periodic coverage of the Poincaré sphere within regions where $\alpha$ and $\delta$ sweep a range of $\pi$ and $2\pi$, respectively, through an equirectangular projection with the poles located at the $s_1$ axis. In this projection, meridians (parallels) are proportionally mapped to $x$ ($y$) (see \hyperref[sec:maps]{Supplemental Material, Section B}). The resulting complex field is given by $\mathbf{E}= \cos{\tilde{x}} e^{-\im \tilde{y}} \mathbf{x} + \sin \tilde{x} e^{\im \tilde{y}} \mathbf{y}$, and its circular polarization components are
\begin{equation}
		E_{\mathbf{l},\mathbf{r}}=\frac{\cos\left[(\tilde{x}\mp \tilde{y})\right]\mp \im \sin\left[ (\tilde{x}\pm \tilde{y})\right]}{\sqrt2}.
    \label{eq:Square_field_circular_basis}
	\end{equation}
    Figure \ref{fig:square}(a) presents the intensity and phase distributions of $E_{\mathbf{l},\mathbf{r}}$. Each polarization component consists of a square array of optical vortices whose superposition results in a lattice of square merons with a polarization state distribution depicted in Fig.~\ref{fig:square}(b). Note that the unit cell of the lattice is formed by four merons, each being one of the four types in Fig.~\ref{fig:merons}(b). Consequently, $N_\mathrm{S}=0$ within the unit cell.

    Expanding the trigonometric functions in $\mathbf{E}$ into a sum of complex exponential terms reveals that the field is composed of a superposition of four plane waves, with two waves polarized along the $x$ direction and two along the $y$ direction. This decomposition illustrates that $k_\perp$ is indeed the absolute value of the transverse component of wavevectors of the plane waves. The diffraction orders in the field's angular spectrum depicted in Fig.~\ref{fig:square}(b) indicate that the transverse wavevectors of the four plane waves lie on the same ring, so that the texture is propagation-invariant.
    Figure \ref{fig:square}(c) displays the distribution of $\mathbf{s}$. The Skyrme density distribution is given by $\rho_\mathrm{S} =  4 k^2_\perp \sin{2 k_\perp x}$, resulting in oscillations in both sign and value along $\tilde{x}$, as illustrated in Fig.~\ref{fig:square}(d).

    Note that introducing an additional relative phase between $E_{\mathbf{l},\mathbf{r}}$ would result in a local rotation of each polarization ellipse by the same amount. This gives rise to a family of square lattices having the same $\rho_\mathrm{S}$ distribution, allowing for the generation of different types of merons  (such as Neel or Bloch-type, as shown in Fig.~\ref{fig:merons}). This transformation can similarly be applied to the other two categories of lattices introduced in this work.

\subsection{Triangular paraxial optical meron lattices}
\label{sec:triangular_lattice}
\begin{figure}
    \centering
    \def\svgwidth{0.48\textwidth}
    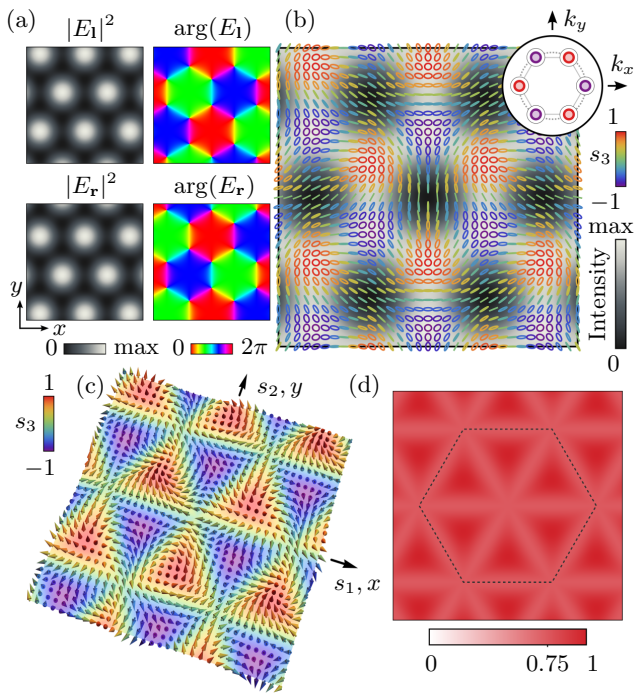\caption{Propagation-invariant triangular optical meron lattice with uniform sign of the Skyrme density. (a) Intensity and phase of the circular polarization components. (b) Polarization ellipse distribution, and angular spectrum of the field (inset) where each dot represents a diffraction order associated with a plane wave whose polarization state is depicted. (c) Normalized Stokes vector distribution and (d) Skyrme density distribution divided by $k^2_\perp$, with its minimum value being 0.75. The unit cell of the texture is enclosed within dashed lines.}
    \label{fig:triangular_texture}
    \end{figure}

    We now introduce a more unconventional family of textures. These are triangular meron lattices (Fig.~\ref{fig:triangular_texture}) that not only remain invariant under propagation, but also exhibit uniform $\mathrm{sgn}(\rho_\mathrm{S})$ across the entire transverse plane. The circular polarization components of a field rendering these lattices are
	\begin{equation}
		E_{\mathbf{l},\mathbf{r}}=e^{\mp \im \tilde{x}} + 2 \; e^{\pm i\tilde{x} / 2} \cos{\left(\frac{\sqrt3}{2}  \tilde{y} \pm \frac{2\pi}{3}\right)}.
		\label{eq:field_hexagon}
	\end{equation}
    Figure \ref{fig:triangular_texture}(a) shows the intensity and phase distributions for $E_{\mathbf{l},\mathbf{r}}$. Each component $E_{\mathbf{l},\mathbf{r}}$ consist of an array of optical vortices forming a hexagonal pattern. Within the same hexagon, the vertices exhibit alternating topological charges. The arrangement of vortices is chosen to ensure that regions with a vortex in one polarization component align with regions lacking a vortex in the other, resulting in merons having  $N_\mathrm{S}=+1/2$. Additionally, there are points where vortices from both circular components with identical topological charges coincide, giving rise to scalar vortices. The zeros in the field emerge as a consequence of fields that maintain a consistent $\mathrm{sgn}(\rho_\mathrm{S})$, and cannot be eliminated \cite{cartography_skyrmions}.

    Figure \ref{fig:triangular_texture}(b) illustrates the field's distributions of polarization ellipses and intensity. The merons take the shape of equilateral triangles, each containing a C-point at its midpoint. The unit cell of the texture is a hexagon, including six merons. Note that the triangular arrangement of the texture's merons, along with their numbers $p$ and $m_\phi$, matches that of a triangular lattice achieved through a conformal transformation introduced in the context of cartography \cite{cartography_skyrmions}, known as Adam's \textit{world in a hexagon projection}  \cite{Adams_hexagon}. The texture presented in this section is not conformal, but it is invariant under propagation, owing to the fact that the transverse wavevectors of the plane waves forming the field align along a ring (with radius $k_\perp$), as depicted in the inset in Fig.~\ref{fig:triangular_texture}(b). Moreover, in theory, propagation does not disturb the zeros of the field.
    
    The angular spectrum of the field is found by expanding Eqs.~\eqref{eq:field_hexagon} into complex exponentials representing individual plane waves. Each component $E_{\mathbf{l},\mathbf{r}}$ is a mix of three plane waves. The entire field is composed of six plane waves, characterized by their transverse wavevectors situated at the vertices of a hexagon. Adjacent vertices of the hexagon belong to orthogonal circular polarization components.
    Figure \ref{fig:triangular_texture}(c) illustrates the distribution of $\mathbf{s}$, and Fig.~\ref{fig:triangular_texture}(d) illustrates $\rho_\mathrm{S}$, which only takes positive values. The fact that $\mathrm{sgn}(\rho_\mathrm{S})>0$ means that all left-handed C-points present lemon patterns, whereas all right-handed C-points exhibit star patterns \cite{cartography_skyrmions}. Within the unit cell of the texture, $N_\mathrm{S}=+3$.
    
    Other members of the same field family (see \hyperref[sec:supplemental_theoretical_results]{Supplemental Material, Section C}) can be explored by applying the transformation $x\rightarrow-x$ to both circular components in Eqs.~\eqref{eq:field_hexagon}. This yields merons with $p=-1/2$, $m_\phi=+1$  (right-handed lemons) and with $p=+1/2$, $m_\phi=-1$ (left-handed stars). Consequently, for every meron, $N_\mathrm{S}=-1/2$, and $\mathrm{sgn}(\rho_\mathrm{S})<0$ throughout space. As mentioned earlier, adjusting the relative phases between $E_{\mathbf{l},\mathbf{r}}$ allows for the exploration of different lattices with the same $\rho_\mathrm{S}$ distribution.
       
    It is worth noting that applying the reflection $x\rightarrow-x$ only to $E_\mathbf{l}$ results in a triangular polarization lattice \cite{lemon_fields} that exhibits oscillations in $\mathrm{sgn}(\rho_\mathrm{S})$, placing it in a distinct family of lattices.
    This transformation also translates into a reflection along the vertical axis in Fourier space affecting the diffraction orders with polarization state $\mathbf{l}$. Consequently, the orders of both $E_{\mathbf{l},\mathbf{r}}$ components coincide in Fourier space at the vertices of an equilateral triangle (so the lattice is also invariant under propagation). This yields a field composed of three linearly polarized waves oriented at 0 and $\pm\pi/3$ rad relative to the $x$ axis. Additional information regarding the skyrmionic properties of this field can be found in \hyperref[sec:lemon_fields]{Supplemental Material, Section A}.
    
\section{Half-Poincaré meron lattices}
\label{sec:half_Poincare}

The final family of fields we present is even more unusual. These fields exhibit square meron lattices where each meron is mapped onto the same Poincaré hemisphere while preserving $\mathrm{sgn}(\rho_\mathrm{S})$. These textures exclusively consist of a single meron type among the four displayed in Fig.~\ref{fig:merons}(b). Moreover, their intensity and $\rho_\mathrm{S}$ distributions remain invariant under propagation. For a field composed only of left-handed ($p=+1/2$) lemon-type ($m_\phi=+1$) merons, the circular polarization components are given by:
\begin{subequations}
\label{eq:half_Poincare_field}
\begin{eqnarray}
		E_{\mathbf{l}}&=&\sin \tilde{x} - \im \sin \tilde{y},
  \\
        E_{\mathbf{r}}&=&\frac{1}{\sqrt2}\left[\sin\left(\tilde{x} -\tilde{y}\right)- \im \sin\left(\tilde{x} +\tilde{y}\right)\right].
  \end{eqnarray}
\end{subequations}

\begin{figure}
    \centering
    \def\svgwidth{0.48\textwidth}
    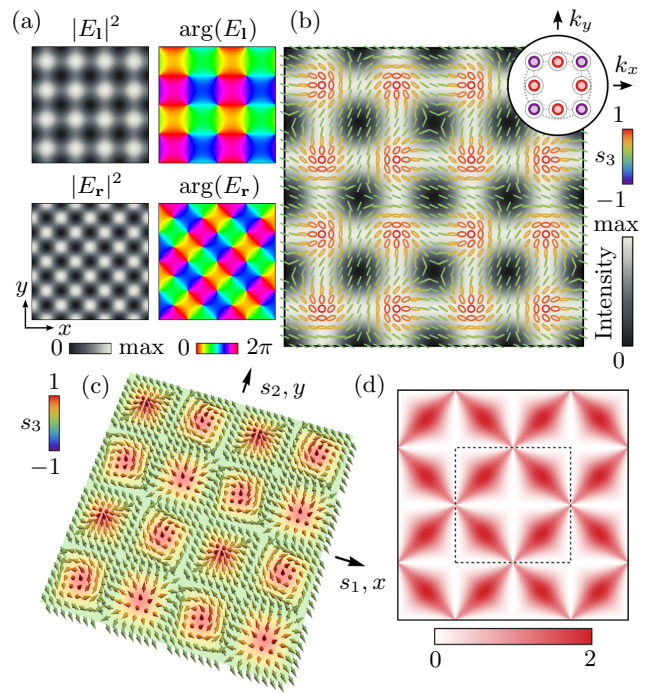\caption{Half-Poincaré optical meron lattices with uniform $\mathrm{sgn}(\rho_\mathrm{S})$. The texture preserves its topology under propagation. (a) Intensity and phase of the circular polarization components. (b) Polarization ellipse distribution, and angular spectrum of the field (inset) where each dot represents a diffraction order associated with a plane wave whose polarization state is depicted. (c) Normalized Stokes vector distribution and (d) Skyrme density distribution divided by $k^2_\perp$. The unit cell of the texture is enclosed within dashed lines.}
    \label{fig:half_Poincare}
    \end{figure}

The intensity and phase distributions of $E_{\mathbf{l},\mathbf{r}}$ are depicted in Fig.~\ref{fig:half_Poincare}(a), while their polarization and intensity distributions are shown in Fig.~\ref{fig:half_Poincare}(b). The unit cell of the texture consist of the four central merons illustrated in \ref{fig:half_Poincare}(b). The spatial distributions in $E_{\mathbf{l},\mathbf{r}}$ are essentially the same as those forming the square lattice in Subsection \ref{sec:square_field}. However, $E_{\mathbf{r}}$ now oscillates spatially at a rate $\sqrt2$ times faster than $E_{\mathbf{l}}$, and the distributions $E_{\mathbf{l},\mathbf{r}}$ are rotated by $\pi/4$ rad with respect to each other. These specific choices ensure that, in the zones where a vortex in a circular polarization component coincides with a region without a vortex in the orthogonal component, vortices always belong to the same circular component and possess the same topological charge. Consequently, the resulting meron always exhibits the same numbers $p$ and $m_\phi$. Additionally, scalar vortices emerge at points where two vortices with the same charge in both circular polarization components coincide, while V-points arise when the vortices have equal charges but opposite signs.

The angular spectrum of the field is illustrated in the inset in Fig.~\ref{fig:half_Poincare}(b), where each diffraction order corresponds to a plane wave composing the field. Each circular polarization component is formed by a superposition of four plane waves whose transverse wavevectors are located at the edges of a square. These square arrangements are rotated by $\pi/4$ rad with respect to one another. 

The diffraction orders in the angular spectrum associated with each circular polarization component are arranged along a ring. Consequently, the transverse profile of each component remains invariant under propagation. However, the radius of this ring for each polarization component differs ($k_\perp$ for $E_{\mathbf{l}}$ and $\sqrt2 k_\perp$ for $E_{\mathbf{r}}$), causing a relative phase shift between them 
under propagation, which results in the the rotation of each polarization ellipse as the field propagates. Using the paraxial approximation, the rotation angle is found to be given by $-k^2_\perp z/(4 k)$, where $k=2\pi/\lambda$, with $\lambda$ being the wavelength. This phenomenon is analogous to what occurs in full Poincaré beams \cite{full_Poincare_beams}, where the polarization ellipse undergoes a rotation of $\pi$ rad locally over the entire propagation distance. In contrast, for the fields we present here, the ellipses continue rotating indefinitely, their angle being proportional to $z$. The vector $\mathbf{s}$ also rotates indefinitely with propagation (at twice the angular velocity as the ellipses), so that the merons oscillate between Neel and Bloch types. 
The evolution of the polarization ellipses and $\mathbf{s}$ vector distributions upon propagation is presented in the \textcolor{blue}{Supplemental Videos 1} and \textcolor{blue}{2}, respectively.
Despite these rotations, the numbers $p$, $m_\phi$, $N_\mathrm{S}$ for each meron, and the Skyrme density distribution, remain unaltered during propagation. 
    
Other square lattice consistenly spanning a single Poincaré hemisphere in the same sense can be attained by modifying Eqs.~\eqref{eq:half_Poincare_field}. Applying the transformation $x\rightarrow-x$ or $y\rightarrow-y$ to $E_{\mathbf{r}}$ yields a field of left-handed stars ($N_\mathrm{S}=-1/2$, $p=+1/2$, $m_\phi=-1$). Alternatively, swapping $E_{\mathbf{l},\mathbf{r}}$ results in a field composed of right-handed stars ($N_\mathrm{S}=+1/2$, $p=-1/2$, $m_\phi=-1$). Following this exchange, one can further apply the transformation $x\rightarrow-x$ or $y\rightarrow-y$ to the current $E_{\mathbf{l}}$ component, which results in a field composed of right-handed lemons ($N_\mathrm{S}=-1/2$, $p=-1/2$, $m_\phi=+1$). For more details of the textures that these fields exhibit, please refer to the \hyperref[sec:supplemental_theoretical_results]{Supplemental Material, Section C}.

Lastly, note that by increasing the amplitude of $E_{\mathbf{r}}$ with respect to $E_{\mathbf{l}}$ in Eqs.~\eqref{eq:half_Poincare_field}, the confinement region of the merons shrinks. This results in smaller, localized areas with high $\rho_\mathrm{S}$ where the merons are located. These regions are surrounded by areas where $\rho_\mathrm{S}$ is positive but nearly zero. In these surrounding regions, the (nearly linear) polarization ellipses have the opposite handedness than within the merons, and the spatial polarization state variation is slow. As $E_{\mathbf{r}}$ is increased, the merons become more confined, but the intensity within them decreases. By further augmenting the amplitude of $E_\mathbf{r}$, one could eventually achieve a lattice of superoscillatory merons (each confined within a very low-intensity area), i.e., merons occupying a region of spatial dimensions  smaller than the wavelength. See \hyperref[sec:supplemental_theoretical_results]{Supplemental Material, Section C} for further information on these textures. It is important to note that all the variations of the lattices presented in this section share the same properties under propagation.

\section{Experimental implementation}

In practice, it is impossible to implement discrete summations of plane waves since they would extend infinitely in space. Therefore, one must implement a spatially-limited version of these fields. Consequently, invariance is only approximate and limited to a fixed distance. 
We implement experimentally spatially-limited versions of the distributions $E_{\mathbf{l},\mathbf{r}}$ presented in Eqs.~\eqref{eq:Square_field_circular_basis}, \eqref{eq:field_hexagon}, and \eqref{eq:half_Poincare_field} in the circular polarization components of a continuous-wave laser beam (with a wavelength of $\lambda=532$~nm) by using a spatial light modulator (SLM). The setup involves dividing the beam using a Wollaston prism, allowing each polarization component to be independently modulated on separate halves of the SLM screen \cite{vector_setup}. We employ a phase-only SLM to modulate both the amplitude and phase of each component with a well-established technique \cite{method_modulation}. We display a circular region on the SLM screen for each circular component containing several spatial periods of the lattices. After the modulation process, the split beam is recombined using the same Wollaston prism. A telescopic system is then employed to image the SLM screen onto a camera. To compute the textures, we make use of intensity measurements obtained from various polarization projections. Additional information regarding the experiment can be found in \hyperref[sec:experimental_setup]{Supplemental Material, Section D}.

\begin{figure}
    \centering
    \def\svgwidth{0.49\textwidth}
    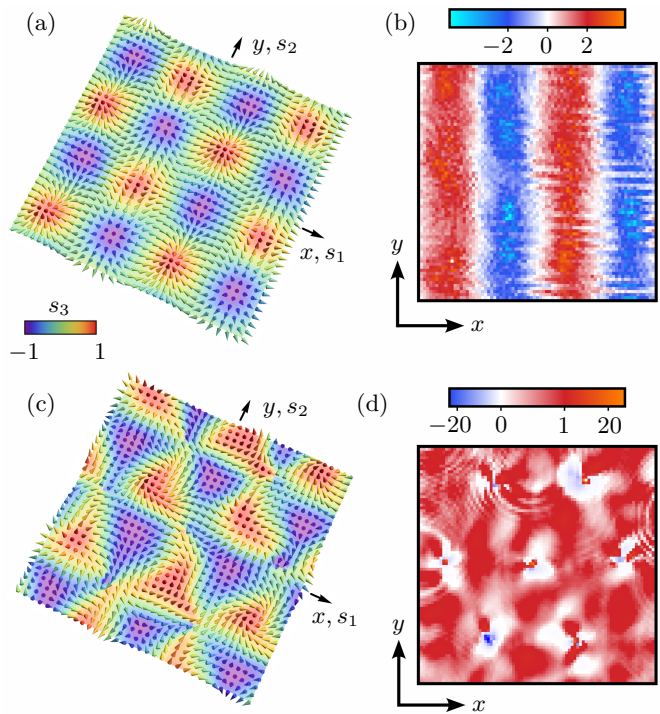\caption{Experimental (a,c) Stokes vector distributions and (b,d) Skyrme density divided by $k^2_\perp$ for the propagation-invariant (a,b) square and (c,d) triangular lattices. The color map for the Skyrme density within the expected theoretical values is consistent with the maps in Figs.~\ref{fig:square}(d) and \ref{fig:triangular_texture}(d). Experimental values exceeding the maximum expected theoretical values are depicted by extending the color map towards orange tones beyond the corresponding theoretical limits (4 and 1 for the square and triangular texture, respectively). Points falling below the theoretical minimum value for the square texture (-4) are represented by a color gradient towards cyan tones. Note that the scale in the Skyrme density color bar for the triangular texture between 0 and 1 increments at a faster rate than the scale for negative values and for values greater than 1; the scale is linear in each of these intervals.}
  \label{fig:experiments_prop_invariant}
    \end{figure}

\begin{figure*}
    \centering
    \def\svgwidth{0.999\textwidth}
    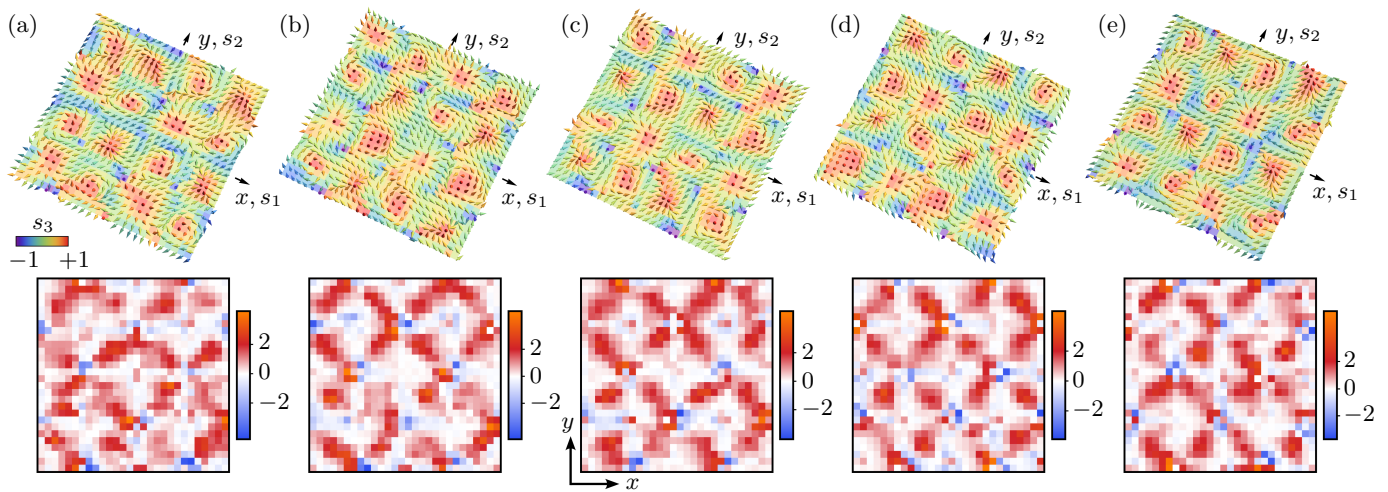\caption{Experimental implementation of half-Poincaré meron lattices. The figure shows the Stokes vector (up) and the Skyrme density distributions divided by $k^2_\perp$ (down) for different transverse planes of propagation. The Stokes vector distribution undergoes a local periodic rotation as the field propagates, whereas the Skyrme density ideally remains unaltered, and experiences only slight changes in practice.  Experimental values in the Skyrme density distribution that surpass the maximum expected theoretical value (2) are represented by extending the color map towards orange tones beyond the corresponding theoretical limit.}
    \label{fig:experiments_propagation}
    \end{figure*}

Figure~\ref{fig:experiments_prop_invariant} show the experimental results for the square (a,b) and triangular (c,d) propagation-invariant lattices, in particular their distributions in $\mathbf{s}$ (a,c) and $\rho_\mathrm{S}$ (b,d). In Fig.~\ref{fig:experiments_propagation}, we present the measured $\mathbf{s}$ and $\rho_\mathrm{S}$ distributions at different planes of propagation for the square lattice where each meron maps to the same Poincaré sphere hemisphere. At each point, the vector $\mathbf{s}$ experiences a negative $2\pi$ rad rotation within the selected distance as the fields propagates towards increasing positive $z$ values, equivalent to a negative $\pi$ rad rotation of the polarization ellipse. The measured polarization ellipse distributions for all fields can be found in \hyperref[sec:supplemental_experimental_results]{Supplemental Material, Section E}. Note that the distributions of $\rho_\mathrm{S}$ exhibit minimal changes as the field propagates.

The transverse planes in Fig.~\ref{fig:experiments_propagation} were captured by displacing the camera's position up to two positions before and after the focal plane of the last lens of the telescopic system, which corresponds to the plane depicted in Fig.~\ref{fig:experiments_propagation}(c). The theoretical  distribution for $\mathbf{s}$ depicted in Fig.~\ref{fig:half_Poincare}(c) is achieved in the plane shown in Fig.~\ref{fig:experiments_propagation}(d).

Note that the resolution of the measurements in Fig.~\ref{fig:experiments_propagation} is lower than that of Fig.~\ref{fig:experiments_prop_invariant}. This difference arises from the telescopic system used to image the field in Fig.~\ref{fig:experiments_propagation}, which was built to produce smaller beam profiles than those in Fig.~\ref{fig:experiments_prop_invariant}. This adjustment was made to achieve a full $2\pi$ rad rotation of the $\mathbf{s}$ vectors within a shorter distance, which is more convenient for experimental purposes. 
The spatial period of the polarization pattern in Fig.~\ref{fig:experiments_propagation} is 72.8~{\textmu}m, corresponding to a value of $2\pi/($ 72.8~{\textmu}m) for $k_\perp$. 
Given that a $4\pi k k^{-2}_\perp$ increase in $z$ is necessary for $\mathbf{s}$ vectors to complete a local $2\pi$ rad rotation, the $\mathbf{s}$ distribution then repeats after a 2~cm displacement in $z$. Nevertheless, increasing the transverse spatial period of the polarization pattern, i.e., decreasing $k_\perp$, allows the creation of fields with an arbitrarily slow variation of the polarization state along $z$. This results in highly stable polarization patterns under $z$ displacements, albeit at the expense of enlarging the unit cell of the polarization state distribution. For example, if the spatial period of the polarization pattern were the same as that for the square lattice in Fig.~\ref{fig:experiments_prop_invariant}(a,b) (197.6~{\textmu}m), a 14.7 cm displacement in $z$ would be required for the $\mathbf{s}$ vectors to complete a local $2\pi$ rad rotation.

Notice that small areas of significantly negative $\rho_\mathrm{S}$ appear for both triangular and half-Poincaré lattices (Figs.~\ref{fig:experiments_prop_invariant}(d) and \ref{fig:experiments_propagation}) near the points where Eqs.~\eqref{eq:field_hexagon} and \eqref{eq:half_Poincare_field} predict field zeros. This phenomenon arises due to the instability of these zeros; even a slight misalignment between the circular components leads to the formation of pairs of C-points with opposite handedness emanating from these singular vertices. In these areas, the Poincaré sphere is wrapped in the opposite direction within a small region of low intensity. 
As pointed out earlier, paraxial fields presenting periodic skyrmionic textures with uniform $\mathrm{sgn}(\rho_\mathrm{S})$ inevitably present such unstable zeros \cite{cartography_skyrmions}. However, the preservation of the skyrmionic structure under propagation suggests that the zeros in the fields presented here are comparatively more stable than those in other textures with uniform $\mathrm{sgn}(\rho_\mathrm{S})$ that are destroyed upon propagation, such as textures based on conformal cartographic projections \cite{cartography_skyrmions}.
Furthermore, the limited plane-wave spectrum of these fields enhances their robustness compared to those based on cartographic maps, as the latter exhibit a more complex spectra. In fact, regions with opposite $\mathrm{sgn}(\rho_\mathrm{S})$ already appear when higher diffraction orders in the angular spectrum of the fields based on cartographic mappings are clipped by the aperture.

We calculated estimates of the Skyrme density based on the pixelated experimental data. Given the small regions of low intensity and negative $\rho_\mathrm{S}$ caused by instabilities as mentioned above, these values differ from the theoretical ones. Within the unit cell of the triangular texture, we measured a value of 2.61 for the Skyrme number, whereas theoretically, it is expected to be 3. In the case of half-Poincaré lattices, $N_\mathrm{S}$ is anticipated to be 2 within the unit cell, while we obtained experimental values of 2.11, 1.67, 1.76, 1.56, and 1.99, for the central unit cell in Fig.~\ref{fig:experiments_propagation}(a-e), respectively.

Within the unit cell of the square lattice with oscillating $\mathrm{sgn}(\rho_\mathrm{S})$ in Fig.~\ref{fig:experiments_prop_invariant}(a,b), the Skyrme number was measured to be -0.002. This result is closer to the expected value of $N_\mathrm{S}=0$ because of the absence of zeros in the field causing these regions of high negative values of $\rho_\mathrm{S}$.

\section{conclusions}

We presented paraxial optical fields having a simple angular spectrum, exhibiting lattices of merons in their polarization state distribution at every plane transverse to the direction of propagation. Two of the proposed families of lattices are invariant under propagation and consist of square and triangular merons, the former characterized by uniform intensity and the latter by a uniform $\mathrm{sgn}(\rho_\mathrm{S})$. Additionally, we introduced more exotic square meron lattices in which all merons map to the same Poincaré hemisphere in the same sense, thereby preserving $\mathrm{sgn}(\rho_\mathrm{S})$. In these lattices, the polarization ellipse undergoes local rotation during propagation periodically. The velocity of this rotation can be reduced by increasing the dimensions of the unit cell of the pattern, making the field more paraxial.

It would be interesting to know what field has the simplest Fourier spectrum that results in a lattice of square merons that covers periodically the entire Poincaré sphere with uniform $\mathrm{sgn}(\rho_\mathrm{S})$. While recent propositions have introduced square lattices spanning the Poincaré sphere using a conformal cartographic map, known as Peirce's \textit{quincuncial projection} \cite{Peirce}, these structures possess a complex spectrum and lack propagation invariance \cite{cartography_skyrmions}. With the purpose of finding a propagation-invariant square lattice spanning the Poincaré sphere with uniform $\mathrm{sgn}(\rho_\mathrm{S})$, one can explore combinations of plane waves to generate configurations of vortices arranged in a square pattern in each circular polarization component, as those in the square lattice in Subsection \ref{sec:square_field}. Nevertheless, it is not hard to realize that one cannot achieve square lattices maintaining $\mathrm{sgn}(\rho_\mathrm{S})$ by aligning displaced, rotated or reflected versions of such patterns in the circular polarization components. Our hypothesis is that propagation-invariant square lattices spanning the Poincaré sphere while preserving $\mathrm{sgn}(\rho_\mathrm{S})$ are not achievable.

\begin{acknowledgments}
This research received funding from the Agence Nationale de Recherche (ANR) through the project 3DPol, ANR-21-CE24-0014-01. D. M. acknowledges Ministerio de Universidades, Spain, Universidad Miguel Hernández and the European Union (Next generation EU fund) for a Margarita Salas grant
from the program Ayudas para la Recualificación del Sistema Universitario Español.
\end{acknowledgments}
 
\bibliography{apssamp}

\clearpage

\section*{Supplemental Information}

\subsection{Triangular paraxial meron lattice with oscillating sign of the Skyrme density}\label{sec:lemon_fields}

\begin{figure}
    \centering
    \def\svgwidth{0.48\textwidth}
    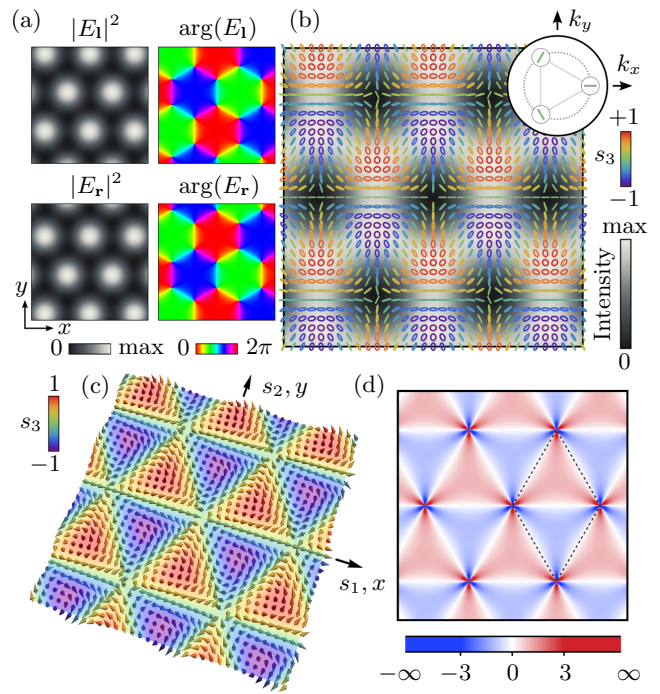\caption{Triangular lattice with oscillating Skyrme density sign \cite{lemon_fields}. (a) Intensity and phase of the circular polarization components. (b) Polarization ellipse distribution and angular spectrum of the field (inset) where each dot represents a diffraction order associated with a plane wave. The polarization state for each diffraction order is depicted in the inset. (c) Normalized Stokes vector and (d) Skyrme density distributions. The unit cell of the texture is enclosed within dashed lines. Note that the unit cell can be rearranged into a rectangle. The Skyrme density tends to $\pm \infty$ close to the fields's zeros. The color map in the Skyrme density remains constant for values beyond $\pm3$.}
    \label{fig:lemon_fields}
    \end{figure}

In Subsection \ref{sec:triangular_lattice} of the main text, we discussed that a triangular meron lattice maintaining $\mathrm{sgn}(\rho_\mathrm{S})$ can be transformed into a distinct triangular lattice with oscillating $\mathrm{sgn}(\rho_\mathrm{S})$ \cite{lemon_fields}. This second field is achieved by applying the transformation $x\rightarrow-x$ to $E_\mathbf{l}$ in Eq.~\eqref{eq:field_hexagon} in the main text. Here, we present an analysis of the skyrmionic  properties of this field. Figure~\ref{fig:lemon_fields}(a) illustrates the intensity and phase distributions of $E_{\mathbf{l},\mathbf{r}}$, while Fig.~\ref{fig:lemon_fields}(b) displays the polarization ellipse distribution and the field angular spectrum. Figures \ref{fig:lemon_fields}(c,d) depict the distributions of $\mathbf{s}$ and $\rho_\mathrm{S}$.

The transformation $x\rightarrow-x$ applied to $E_\mathbf{l}$ reverses the sign of the topological charges of all the vortices in $\phi_\mathbf{l}$, leading to a change in the sign of $m_\phi$ within all the right-handed merons without affecting $p$. Consequently, all right-handed ($p=-1/2$) merons, previously being stars ($m_\phi=-1$) before the transformation, now exhibit lemon-type patterns ($m_\phi=+1$). This leads to a change in sign of $N_\mathrm{S}$ within the right-handed merons. The lattice now exclusively consists of right- and left-handed lemon-type C-points, resulting in oscillations in $\mathrm{sgn}(\rho_\mathrm{S})$. Additionally, the scalar vortices transform into V-points with $m_\phi=2$.

As outlined in the main text, the transformation $x\rightarrow-x$ corresponds to a reflection of the diffraction orders along the vertical axis in Fourier space. As a consequence, diffraction orders corresponding to both circular components now coincide, forming an equilateral triangle in Fourier space. This field is a mixture of three plane waves with linear polarization states oriented at $0$ and $\pm\pi/3$ rad with respect to the $x$ axis \cite{lemon_fields}.
	
The unit cell of the texture consists of two triangular merons, with each meron mapping onto a different hemisphere of the Poincaré sphere. Within the left (right)-handed meron, $\rho_\mathrm{S}$ is positive (negative), giving rise to $N_\mathrm{S}=\pm1/2$, respectively. Figure \ref{fig:lemon_fields}(d) shows that the Skyrme density tends towards $\pm\infty$ at the field zeros.

Through the substitution $x\rightarrow-x$ in both circular polarization components of the field discussed here, a lattice composed of left and right-handed star-type merons is achieved \cite{stars_fields}.

\subsection{Spherical projections of the textures}\label{sec:maps}

\begin{figure}
    \centering
    \def\svgwidth{0.49\textwidth}
    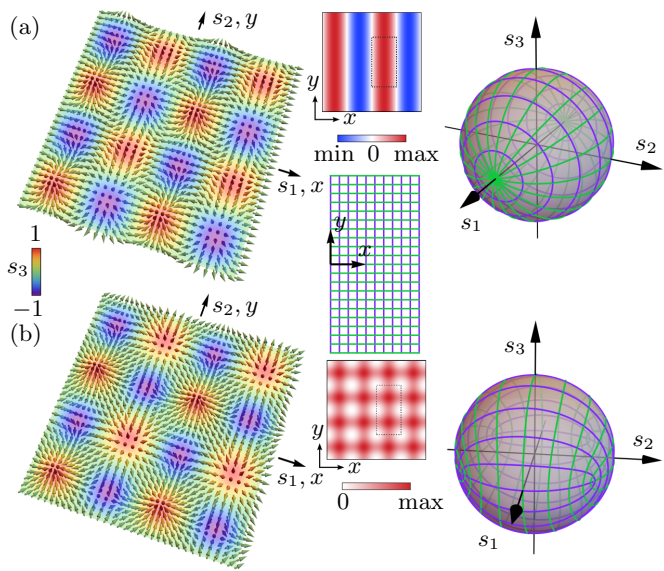\caption{Map of a square grid within a region in physical space onto the Poincaré sphere for a square lattice based on (a) a nonconformal map with variant $\mathrm{sgn}(\rho_\mathrm{S})$ and (b) a conformal map with uniform $\mathrm{sgn}(\rho_\mathrm{S})$. The region mapped is depicted within dashed lines in the Skyrme density distribution.}    \label{fig:square_maps}
\end{figure}

\begin{figure}
    \centering
    \def\svgwidth{0.49\textwidth}
    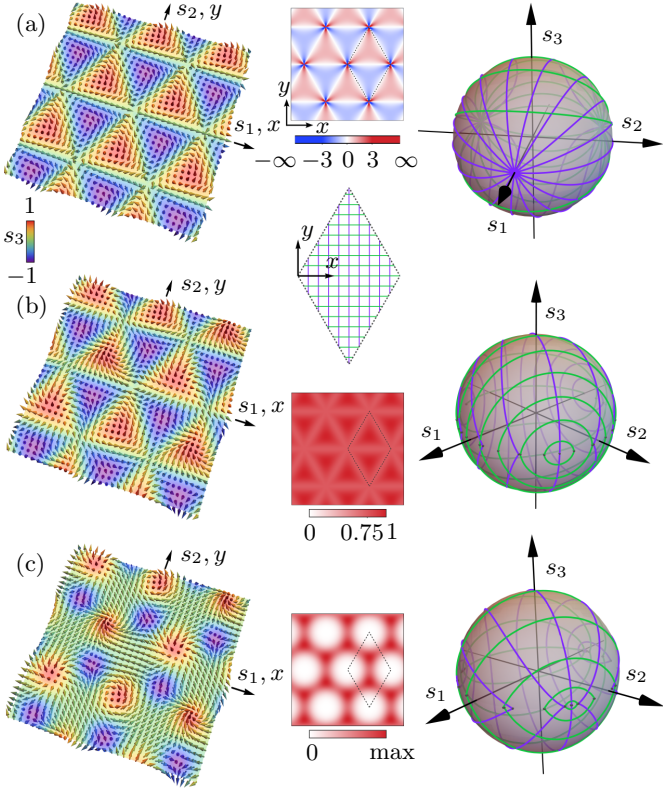\caption{Map of a square grid within a region in physical space onto the Poincaré sphere for a triangular lattice based on a (a) nonconformal map with variant $\mathrm{sgn}(\rho_\mathrm{S})$, (b) nonconformal map with uniform $\mathrm{sgn}(\rho_\mathrm{S})$, and (c) conformal map preserving $\mathrm{sgn}(\rho_\mathrm{S})$. The region mapped is depicted within dashed lines in the Skyrme density distribution.}
    \label{fig:sphere_maps_triangle}
\end{figure}

\begin{figure}
    \centering
    \def\svgwidth{0.49\textwidth}
    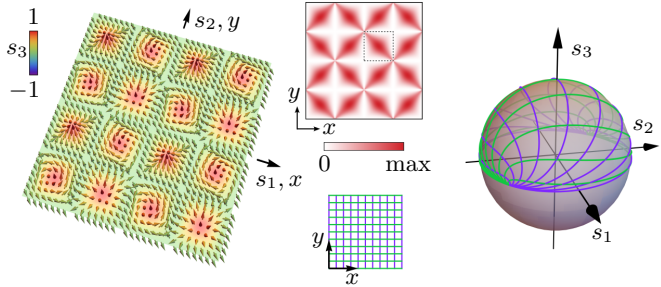\caption{Map of a square grid within a region in physical space onto the Poincaré sphere for a square lattice that always map to the same hemisphere. The texture is nonconformal and preserves $\mathrm{sgn}(\rho_\mathrm{S})$. The region mapped is depicted within dashed lines in the Skyrme density distribution.}
    \label{fig:sphere_map_half_Poincare}
\end{figure}

	Figures \ref{fig:square_maps}, \ref{fig:sphere_maps_triangle}, and \ref{fig:sphere_map_half_Poincare} offer a visualization of how a rectangular grid within a polygon in physical space is projected onto either the entire Poincaré sphere or a single hemisphere. 
 
 In Fig.~\ref{fig:square_maps} we observe the mapping of a rectangular region in physical space onto the full sphere for two distinct lattices of square merons. The square lattice with oscillating $\mathrm{sgn}(\rho_\mathrm{S})$ described by Eqs.~\eqref{eq:Square_field_circular_basis} in the main manuscript is shown in Fig.~\ref{fig:square_maps}(a). The Skyrme density vanishes along a set of equidistant vertical lines that corresponds to the two intersections of the sphere with the $s_1$ axis. These intersections correspond to either horizontally or vertically linearly polarized light and act as singularities in the mapping.

    Let us compare this map with the one in Fig.~\ref{fig:square_maps}(b), which illustrates how a square grid, contained in a rectangle spanning the Poincaré sphere, maps onto the sphere for a texture based on Peirce's \textit{quincuncial projection}. This is a conformal cartographic projection that creates periodic textures preserving $\mathrm{sgn}(\rho_\mathrm{S})$ \cite{cartography_skyrmions}. In this case, $\rho_\mathrm{S}$ is zero at specific points that correspond to the singular points of the transformation, which are four equally spaced points along the equator of the sphere.

    By examining the grid over the sphere for both maps, one can observe that infinitesimally small squares on the plane are mapped to infinitesimally small squares on the sphere in Peirce's (conformal) map, but not in the equirectangular (nonconformal) projection. Additionally, the lines that map onto the same Cartesian planar coordinate $y$ do not intersect at singular points on the sphere in Peirce's map, whereas they do in the equirectangular projection. This is related to the fact that the sign of $\rho_\mathrm{S}$ is not reversed at these points for Peirce's projection, whereas it is reversed in the equirectangular projection.

Figures \ref{fig:sphere_maps_triangle}(a-c) illustrate the mapping process from a planar square grid inside a rhomboidal region to the Poincaré sphere for three distinct lattices of equilateral triangular merons.

In Figure \ref{fig:sphere_maps_triangle}(a), we present the map for the lattice discussed in the previous section in the Supplementary Material \cite{lemon_fields}. The rhomboid contains two merons that, in this case, constitute a unit cell of the texture. Each edge of a triangular meron corresponds to a unique singular point of the mapping lying on the equator of the Poincaré sphere. Due to the identical boundaries for both types of merons, there are only three of these points. They represent linearly polarized states at angles of $0$ and $\pm\pi/3$ relative to the $x$ axis. The Skyrme density is zero at the edges of the merons. Lines of the map corresponding to the same Cartesian coordinate in physical space intersect at these points, resulting in variations in $\mathrm{sgn}(\rho_\mathrm{S})$ from one side of the edge to the other.

The sphere map of the triangular meron lattice preserving $\mathrm{sgn}(\rho_\mathrm{S})$, introduced in Subsection \ref{sec:triangular_lattice} in the main text and described by Eqs.~\eqref{eq:field_hexagon}, is depicted in Fig.~\ref{fig:sphere_maps_triangle}(b). In this case, the rhomboidal region corresponds to one of the three rhomboids that make up the unit cell of the texture, forming a hexagon. 

This transformation also exhibits three singular points equidistantly distributed along the equator. These points correspond to linearly polarized light at angles of $\pi/2$ and $\pm\pi/6$ rad (these points could have been chosen to be the same as the polarization states at the singular points of the previous map).  They coincide with the vertices of the triangles in physical space, where the zeros of the field are located. The Skyrme density evaluates (in the L'H\^opital sense) to a well-defined value of 3/4. The points along the curves located at the edges of the rhomboid are depicted in dark gray. These points illustrate that each closed loop on the sphere actually consists of two distinct curves, mapping two disconnected lines of constant $x$ or $y$, positioned opposite each other relative to the center of the rhomboid in physical space.

The lattice in Fig.~\ref{fig:sphere_maps_triangle}(b) is nonconformal, but it consists in a triangular arrangement of merons having the same numbers $p$ and $m_\phi$ than a texture based of a conformal transformation known as Adam's \textit{world in a hexagon} cartographic projection \cite{Adams_hexagon, cartography_skyrmions}. This case is illustrated in Fig.~\ref{fig:sphere_maps_triangle}(c). The singular points of the map are also positioned at the edges of the triangular merons, and are intentionally chosen to match those in the previous sphere map. In this case, these points coincide with locations of zero Skyrme density. At these points, the lines belonging to the same Cartesian component on the plane present avoided crossings, as occurs in the map in Fig.~\ref{fig:sphere_maps_triangle}(b).

The final transformation that we illustrate corresponds to the textures mapping a single hemisphere, introduced in Section \ref{sec:half_Poincare} and described by Eqs.~\eqref{eq:half_Poincare_field}. Figure \ref{fig:sphere_map_half_Poincare} demonstrates how a square grid within a square region in physical space is projected onto the hemisphere. This region corresponds to one of the square merons that constitute the unit cell of the texture. The resulting map features two singular points that lie on the equator of the sphere. These points correspond to two orthogonal linear polarizations that rotate as the field propagates. They coincide with the two vertices of the square merons where the field's scalar zeros are located. The Skyrme density is zero at these points.

\subsection{Other members of the families of fields} \label{sec:supplemental_theoretical_results}

Here, we present additional examples of fields belonging to the same families as those introduced in the main text.

\begin{figure}
    \centering
    \def\svgwidth{0.49\textwidth}
    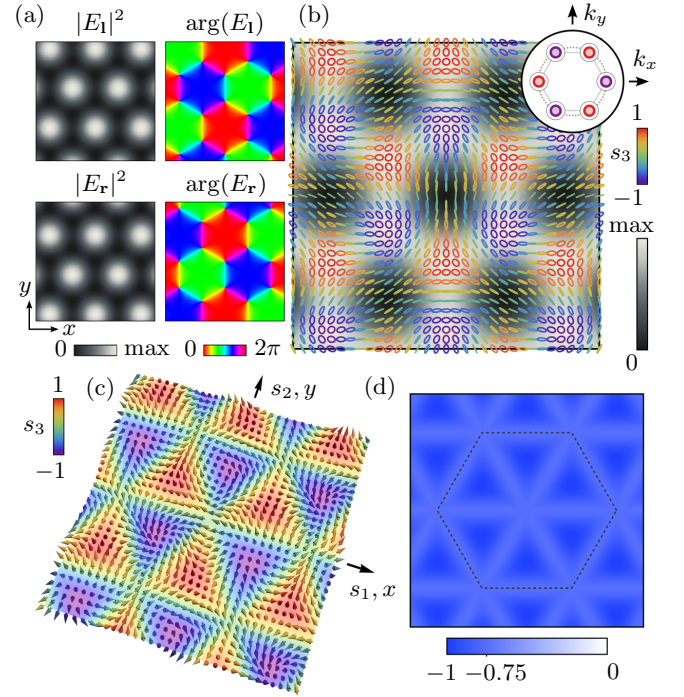\caption{Triangular meron lattice with $\mathrm{sgn}(\rho_\mathrm{S})<0$. (a) Intensity and phase of the circular polarization components. (b) Polarization ellipse distribution and angular spectrum of the field. Each dot stands for a diffraction order associated with a plane wave. The polarization state for each diffraction order is depicted. (c) Normalized Stokes vector and (d) Skyrme density distributions. The unit cell of the texture is enclosed within dashed lines.}
    \label{fig:triangular_lattice_negative_density}
\end{figure}

The triangular lattice formed by left-handed lemons and right-handed stars in Subsection \ref{sec:triangular_lattice} in the main manuscript (with $\mathrm{sgn}(\rho_\mathrm{S})>0$) can be transformed into a lattice of right-handed lemons and left-handed stars by applying the transformation $x\rightarrow-x$ to Eqs.~\eqref{eq:field_hexagon} in the main text. Figure \ref{fig:triangular_lattice_negative_density}(a) depicts the intensity and phase distributions of the circular polarization field components, while Fig.~\ref{fig:triangular_lattice_negative_density}(b) illustrates the polarization ellipse map and the field angular spectrum. Figures \ref{fig:triangular_lattice_negative_density} (c,d) display the distribution of $\mathbf{s}$ and $\mathrm{sgn}(\rho_\mathrm{S})$. It is worth noting that in this configuration, the Skyrme density is negative throughout the entire field.

\begin{figure}
    \centering
    \def\svgwidth{0.49\textwidth}
    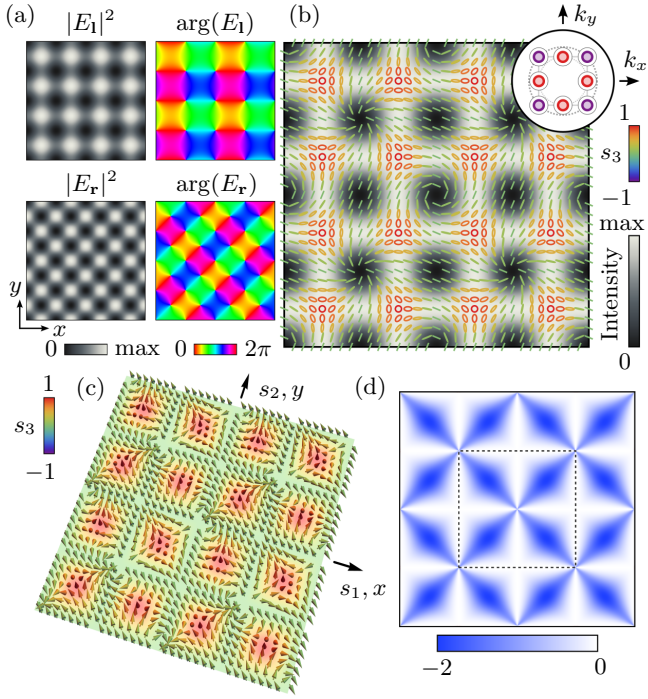\caption{Half-Poincaré meron lattice mapping the northern Poincaré sphere hemisphere with $\rho_\mathrm{S}<0$. (a) Intensity and phase of the circular polarization components. (b) Polarization ellipse distribution and angular spectrum of the field. Each dot stands for a diffraction order associated with a plane wave. The polarization state for each diffraction order is depicted. (c) Normalized Stokes vector and (d) Skyrme density distributions. The unit cell of the texture is enclosed within dashed lines.}
    \label{fig:half_Poincare_lattice_redstars}
\end{figure}

\begin{figure}
    \centering
    \def\svgwidth{0.49\textwidth}
    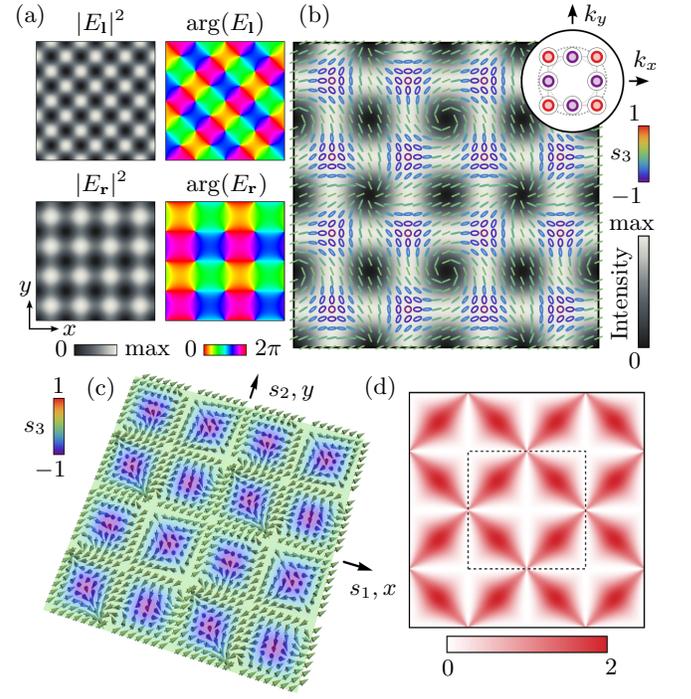\caption{Half-Poincaré meron lattice mapping the southern Poincaré sphere hemisphere with $\rho_\mathrm{S}>0$. (a) Intensity and phase of the circular polarization components. (b) Polarization ellipse distribution and angular spectrum of the field. Each dot represents a diffraction order associated with a plane wave. The polarization state for each diffraction order is depicted. (c) Normalized Stokes vector and (d) Skyrme density distributions. The unit cell of the texture is enclosed within dashed lines.}
    \label{fig:half_Poincare_lattice_bluestars}
\end{figure}

\begin{figure}
    \centering
    \def\svgwidth{0.49\textwidth}
    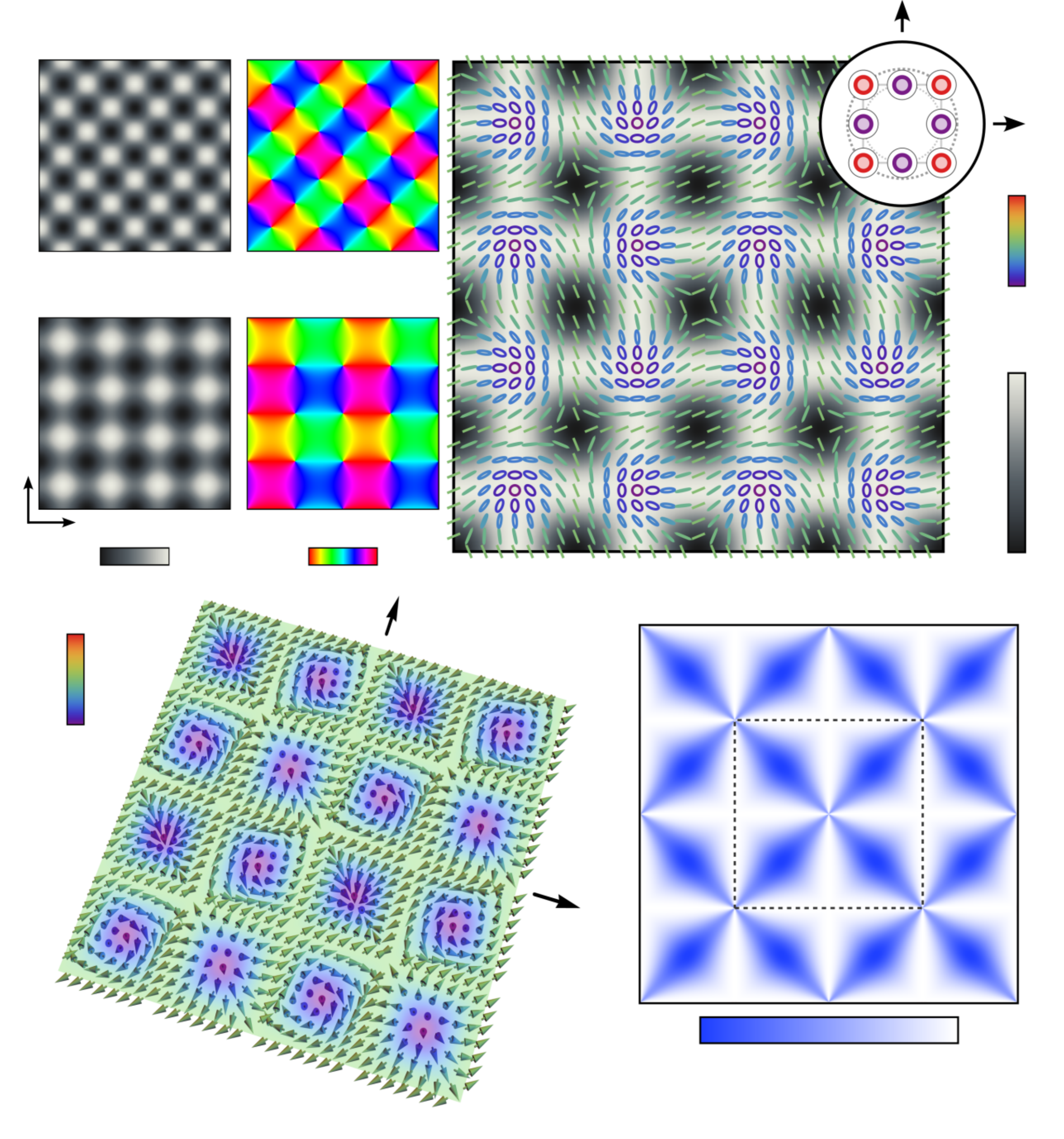\caption{Half-Poincaré meron lattice mapping the southern Poincaré sphere hemisphere with $\rho_\mathrm{S}<0$. (a) Intensity and phase of the circular polarization components. (b) Polarization ellipse distribution and angular spectrum of the field. Each dot represents a diffraction order associated with a plane wave. The polarization state for each diffraction order is depicted. (c) Normalized Stokes vector and (d) Skyrme density distributions. The unit cell of the texture is enclosed within dashed lines.}
    \label{fig:half_Poincare_lattice_bluelemons}
\end{figure}

Similar variations can be explored for the lattices exclusively composed of left-handed lemon-type merons, as introduced in Section \ref{sec:half_Poincare} of the main text. The distribution can be transformed into a lattice consisting only of left-handed stars by setting $x\rightarrow-x$ in $E_\mathbf{r}$ in Eq.~\eqref{eq:half_Poincare_field}. The resulting field and the properties of the texture are illustrated in Fig.~\ref{fig:half_Poincare_lattice_redstars}. The Skyrme density now is negative at every point in space.
     If instead we swap the circular polarization components in Eqs.~\eqref{eq:half_Poincare_field}, we obtain a field consisting of right-handed stars (Fig.~\ref{fig:half_Poincare_lattice_bluestars}). Performing a substitution of the form $x\rightarrow-x$  to the current $E_{\mathbf{l}}$ component after exchanging the circular polarization components results in a texture of right-handed lemons (see Fig.~\ref{fig:half_Poincare_lattice_bluelemons}).

\begin{figure}
    \centering
    \def\svgwidth{0.49\textwidth}
    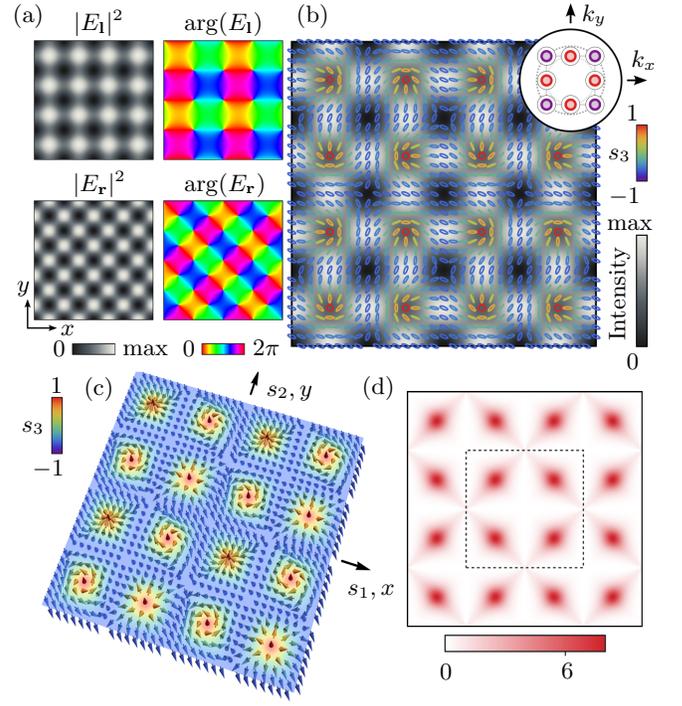\caption{Meron lattice of isolated left-handed lemon-type merons surrounded by right-handed elliptically polarized states. (a) Intensity and phase of the circular polarization components. (b) Polarization ellipse distribution and angular spectrum of the field. Each dot represents a diffraction order associated with a plane wave. The polarization state for each diffraction order is depicted. (c) Normalized Stokes vector and (d) Skyrme density distributions. The unit cell of the texture is enclosed within dashed lines.}
    \label{fig:half_Poincare_isolated}
\end{figure}

It is  mentioned in Section \ref{sec:half_Poincare} of the main manuscript that when the amplitude of $E_\mathbf{r}$ is increased relative to that of $E_\mathbf{l}$ in Eqs.~\eqref{eq:half_Poincare_field}, the regions containing the merons contract, as depicted in Fig.~\ref{fig:half_Poincare_isolated}, and the intensity in these regions diminishes while the Skyrme density increases.  These regions are then surrounded by areas with small negative Skyrme density, where the ellipticity varies slowly and has the opposite handedness of that within the merons.

\clearpage

\subsection{Experimental setup}
\label{sec:experimental_setup}

\begin{figure}
    \centering
    \def\svgwidth{0.48\textwidth}
    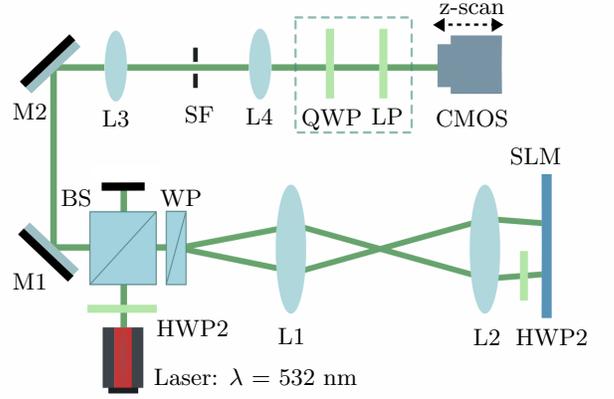\caption{Scheme of the experimental setup employed to generate the meron lattices. The orientation of the linearly polarized state of a continuous-wave (CW) laser beam with a wavelength of $\lambda=532$ nm is controlled using a half-wave plate (HWP1) before entering a 50:50 beamsplitter (BS). The light is then reflected by the BS to a Wollaston prism (WP), which splits it into two orthogonal linearly polarized beams forming an angle of 1$^{\circ}$. Subsequently, both beams undergo magnification through a telescope consisting of lenses L1 and L2 with focal lengths of $f_1 = 100$ mm and $f_2=500$ mm, respectively. After reflecting off the SLM screen, the two components are recombined by the WP. Note that a second half-wave plate is placed in one of the beam paths to convert vertical polarization into horizontal polarization, as the SLM only modulates horizontal polarization. Then, HWP2, upon reflection by the SLM, restores the polarization to its original vertical state. After the WP, the beam returns to the BS, where it is directed to a second telescope imaging the SLM plane onto a camera. The second telescope comprises lenses L3 and L4. The focal lengths of L3 and L4 employed to image the fields described by Eqs.~\eqref{eq:Square_field_circular_basis} and \eqref{eq:field_hexagon} where $f_3 =300 $ mm and $f_4=400 $ mm, respectively. To measure the evolution of the polarization state distribution during the propagation of the field described by Eq.~\eqref{eq:half_Poincare_field}, a different telescope was employed to reduce the size of the lattices. In this case, lenses L3 and L4 had focal lengths of $f_3 =300 $ mm and $f_4=150 $ mm, respectively. Additionally, in this last case, the camera was mounted on a moving stage to capture different planes before and after the SLM's conjugated plane. A spatial filter (SF) was placed at the Fourier plane of the last telescope lens for all measurements. Finally, the polarization state distribution of the lattices is measured using a polarization state analyzer positioned before the camera. This analyzer is composed of a retractable quarter-wave plate (QWP) and a retractable linear polarizer (LP).}
    \label{fig:experimental_setup}
    \end{figure}

The experimental setup for generating the meron lattices is a modified version of the method proposed by Maurer {\it et al}. \cite{vector_setup}. The system is depicted in Fig.~\ref{fig:experimental_setup}, and details are provided in the figure caption.
   
The spatial light modulator (SLM) screen was partitioned into two sections. In each section, the phase and amplitude of each transverse orthogonal polarization component of the field were independently modulated. For simplicity, we modulated the horizontal and vertical polarization components of each field, denoted as $E_{\textbf{x},\textbf{y}}$. These components can be obtained as a linear combination of the circular polarization components, $E_{\textbf{l},\textbf{r}}$, in Eqs.~\eqref{eq:Square_field_circular_basis}, \eqref{eq:field_hexagon}, and \eqref{eq:half_Poincare_field} in the main text, expressed as:
\begin{equation}
    E_{\textbf{x},\textbf{y}}=\frac{c_{\textbf{x},\textbf{y}}}{\sqrt{2}}\left( E_\textbf{r} \pm E_\textbf{l} \right),
  \label{eq:general_field_lineal}
\end{equation}
where \textbf{x}, \textbf{y} are the horizontal and vertical lineal polarization components, and $c_{\textbf{x}}=1$ while $c_{\textbf{y}}=\im$.

The phase and amplitude of each polarization component were modulated in each half of the SLM screen using an algorithm introduced by Bolduc {\it et al}. \cite{method_modulation}. This algorithm encodes amplitude and phase in a phase-only hologram. Assuming that the phase and amplitude of the input field are constant, a function with amplitude $A$ and phase $\Phi$ can be encoded by displaying on the SLM screen a phase function $\Psi=M \mathrm{Mod(} F\mathrm,2\pi{)}$, where $M=1+\frac{1}{\pi}\mathrm{sinc}^{-1}(A)$ and $F=\Phi-\pi M$.

The encoded amplitude is then $A_{\textbf{x},\textbf{y}}=|E_{\textbf{x},\textbf{y}}|$, while the encoded phase is $\Phi_{\textbf{x},\textbf{y}}=\mathrm{Mod}(\mathrm{arg}[E_{\textbf{x},\textbf{y}}]\pm \Phi_{B}+\Phi_{C},2\pi)$. Here, $\Phi_{B}$ is a linear phase displayed on each half of the screen with an opposite sign to ensure the precise recombination of the two polarization components into a Wollaston prism (WP), and $\Phi_{C}$ is a phase provided by the manufacturer to correct the inherent aberrations of the SLM.

The SLM screen is imaged onto a camera. Before the camera, a linear polarizer (LP) and a quarter-wave plate (QWP) are employed to measure the polarization state distribution by capturing the intensity distributions of the linearly polarized horizontal (\textbf{x}), vertical (\textbf{y}), and oriented at $\pm 45^{\circ}$ from \textbf{x} (\textbf{p} and \textbf{m}, respectively) components, as well as the left- and right-circular polarization components (\textbf{l} and \textbf{r}). These intensities are denoted as $I_i$ with $i=\mathbf{x}, \mathbf{y}, \mathbf{p}, \mathbf{m}, \mathbf{l}, \mathbf{r}$. In addition, the total intensity, denoted by $I$, was measured without LP and QWP. We compute the Stokes parameters as:
\begin{equation}
  \begin{aligned}
   & \quad S_0 = I, \\
   & \quad S_1 = I_\textbf{x} - I_\textbf{y}, \\
   & \quad S_2 = I_\textbf{p} - I_\textbf{m},\\
   & \quad S_3 = I_\textbf{l} - I_\textbf{r}.
 \end{aligned}
\end{equation}
The normalized Stokes vector is given by $\textbf{s}=\textbf{s}'/|\textbf{s}'|$, where $\textbf{s}'=(S_1,S_2,S_3)/S_0$. Note that we divide the vector $\mathbf{s'}$ by its norm to avoid values smaller or even greater than 1, which are produced by the limited size of the pixels and noise.

The azimuthal and polar angles on the Poincaré sphere are computed using the components of $\mathbf{s}=(s_1,s_2,s_3)$:
\begin{align}
\phi&=\mathrm{atan2}(s_2,s_1),\\
\theta&=\arccos s_3.
\end{align}
The semi-major and semi-minor axes of the polarization ellipses are
\begin{align}
a&=\left[1+\frac{\tan(\theta-\pi/4)}{2}\right]^{-1/2},\\
b&=\sqrt{1-a^2}.
\end{align}

\subsection{Additional experimental results}
\label{sec:supplemental_experimental_results}

\begin{figure}[h]
    \centering
    \def\svgwidth{0.49\textwidth}
    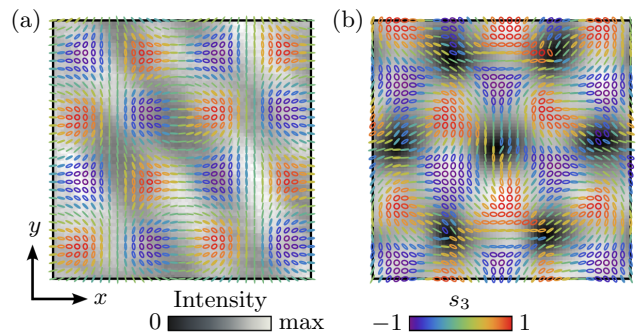\caption{Measured intensity and polarization ellipse distributions for the (a) square and (b) triangular meron lattices in Subsections \ref{sec:square_field} and \ref{sec:triangular_lattice} in the main text, respectively.}    \label{fig:experiments_prop_invariant_ellipses}
\end{figure}

\begin{figure*}[h!]
    \centering
    \def\svgwidth{0.99\textwidth}
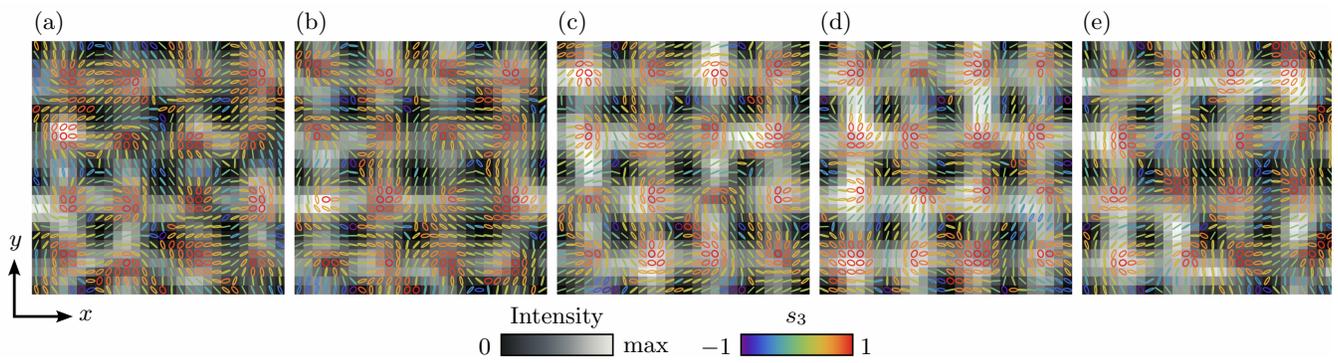\caption{Measured intensity and polarization ellipse distributions during propagation for the half-Poincaré meron lattice in Section \ref{sec:half_Poincare} in the main text.} \label{fig:experiments_propagation_ellipses}
\end{figure*}

Figure \ref{fig:experiments_prop_invariant_ellipses} displays the experimental intensity and polarization ellipse maps for the square and triangular meron lattices in Subsections \ref{sec:square_field} (Fig.~\ref{fig:experiments_prop_invariant_ellipses}(a)) and \ref{sec:triangular_lattice} (Fig.~\ref{fig:experiments_prop_invariant_ellipses}(b)) in the main text.
Figure \ref{fig:experiments_propagation_ellipses} shows the evolution under propagation of the measured intensity and polarization ellipse maps for the half-Poincaré texture in Section \ref{sec:half_Poincare} in the main text.

\end{document}